\documentclass[%
prl,
 reprint,
 amsmath,amssymb,
 aps,preprintnumbers]{revtex4-1}

\usepackage[colorlinks=true,linkcolor=black, citecolor=black,
urlcolor=black]{hyperref}

\usepackage{multirow,graphics}

\usepackage{amstext}
\usepackage{amssymb}
\usepackage{amsmath}
\usepackage{graphicx}
\usepackage{color}

\begin{document}

\newcommand{\IM}{{\rm Im}\,}
\newcommand{\card}{\#}
\newcommand{\la}[1]{\label{#1}}
\newcommand{\eq}[1]{(\ref{#1})}
\newcommand{\figref}[1]{Fig. \ref{#1}}
\newcommand{\abs}[1]{\left|#1\right|}
\newcommand{\comD}[1]{{\color{red}#1\color{black}}}

\makeatletter
     \@ifundefined{usebibtex}{\newcommand{\ifbibtexelse}[2]{#2}} {\newcommand{\ifbibtexelse}[2]{#1}}
\makeatother

\preprint{LPTENS-15/01, DESY 15-086, JLAB-THY-15-2053}

\newcommand{\footnoteab}[2]{\ifbibtexelse{%
\footnotetext{#1}%
\footnotetext{#2}%
\cite{Note1,Note2}%
}{%
\newcommand{\textfootnotea}{#1}%
\newcommand{\textfootnoteab}{#2}%
\cite{thefootnotea,thefootnoteab}}}

\def\e{\epsilon}
     \def\bT{{\bf T}}
    \def\bQ{{\bf Q}}
    \def\wT{{\mathbb{T}}}
    \def\wQ{{\mathbb{Q}}}
    \def\ttQ{{\bar Q}}
    \def\tQ{{\tilde \bP}}
        \def\bP{{\bf P}}
    \def\CF{{\cal F}}
    \def\cC{\CF}
     \def\Tr{\text{Tr}}
     \def\l{\lambda}
\def\hbZ{{\widehat{ Z}}}
\def\bZ{{\resizebox{0.28cm}{0.33cm}{$\hspace{0.03cm}\check {\hspace{-0.03cm}\resizebox{0.14cm}{0.18cm}{$Z$}}$}}}
\newcommand{\rb}{\right)}
\newcommand{\lb}{\left(}

\newcommand{\gT}{T}\newcommand{\gQ}{Q}

\title{Structure constant of twist-2 light-ray operators in Regge limit}

\author{ Ian Balitsky$^{a}$, Vladimir Kazakov$^{b}$, Evgeny Sobko$^{c}$}

\affiliation{%
\(^{a}\)Physics Dept., Old Dominion University, Norfolk VA 23529 \&
 Theory Group, JLAB, 12000 Jefferson Ave, Newport News, VA 23606
\\
             \(^{b}\) LPT, \'Ecole Normale Superieure, 24, rue Lhomond 75005 Paris, France \& Universit\'e Paris-VI, Place Jussieu, 75005 Paris, France\\
 \(^{c}\) DESY Hamburg, Theory Group,
Notkestra{\ss}e 85, 22607 Hamburg, Germany
}

\begin{abstract}
We compute the normalized structure constant of three twist-2 operators   in \(\mathcal{N}=4\) SYM in the leading BFKL approximation at any  \(N_c\). The result is applicable to other gauge theories including  QCD. \end{abstract}

 \maketitle

\section{Introduction}

The problem of high-energy behavior of amplitudes has a long story \cite{Ioffe:2010zz,Kovchegov:2012mbw}. One of the most popular approaches is to reduce the gauge theory at high
  energies to 2+1 effective theory which can be solved exactly or
  by computer simulations. Unfortunately, despite the multitude
  of attempts, the Lagrangian for 2+1 QCD at high energies is not written yet.
  In this context the idea to solve formally the high-energy QCD or \(\mathcal{N}=4\) SYM
  by calculation of anomalous dimensions and structure constants in the BFKL limit seems to be very
  promising.

\(\mathcal{N}=4\) SYM is a superconformal theory and its most important physical properties are encoded   into the OPE  characterized by  the spectrum of anomalous dimensions and by the structure constants. While   the former  is now exactly and efficiently computable  at  large \(N_c\)  due to quantum integrability \cite{Beisert:2010jrGromov:2014caa}, the calculation of the OPE structure constants is these days on a fast track, especially after the ground-breaking    all-loop proposal of \cite{Basso:2015zoa}.

\begin{figure}
  \centering
  \includegraphics[scale=0.235]{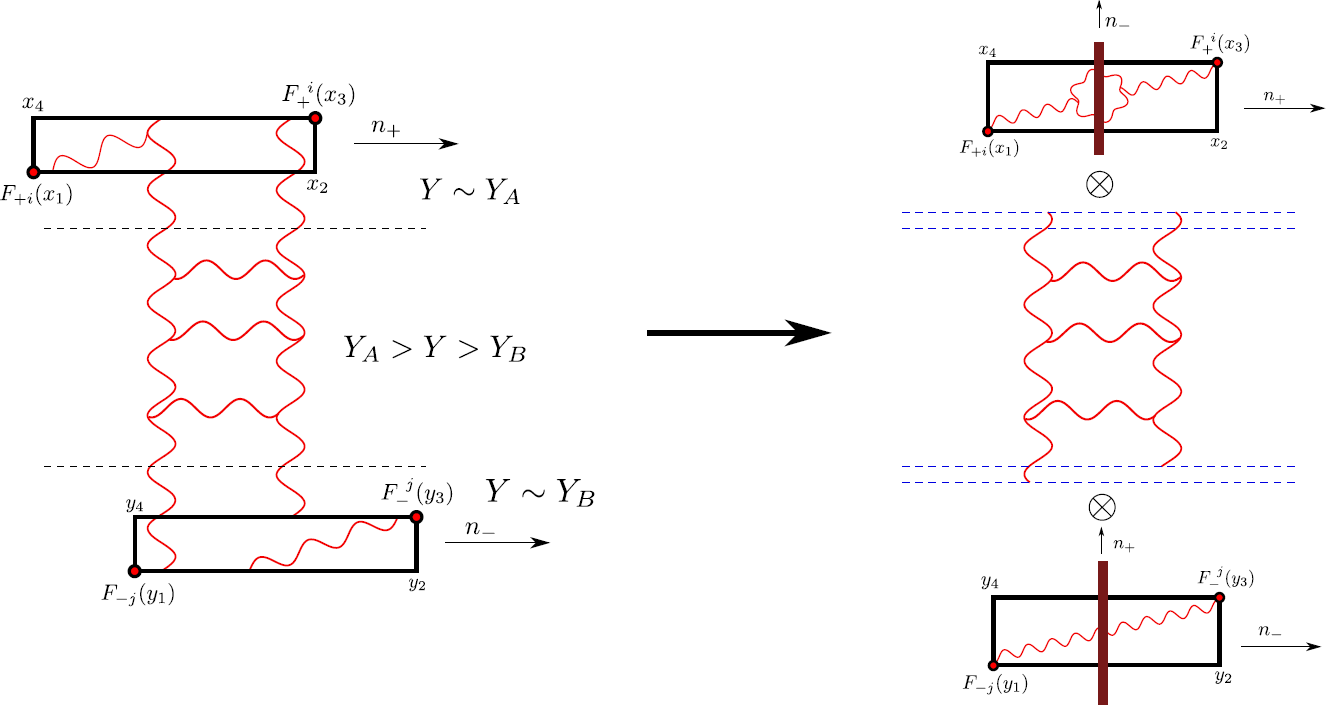}
  \caption{Scheme of computation of 2-point  correlator. In the l.h.s., the long sides of regularizing rectangular Wilson frames are stretched along light ray and the short sides in the orthogonal directions.  In the r.h.s. we use OPE of frames over color dipoles and compute their correlator, see \cite{Balitsky:2013npa} for details.}
  \label{fig:2p}
\end{figure}

In this note we calculate the 3-point correlator of twist-2 operators  \(\mathcal{O}^{j}(x)=\text{tr} F_{+i}D_+^{j-2}F_+^i+fermions+scalars\) in \(\mathcal{N}=4\) SYM in the BFKL limit \cite{Fadin:1975cbBalitsky:1978ic} when \(\omega=j-1\rightarrow 0\), the 't~Hooft coupling \(g^2\equiv \frac{N_c g_{YM}^2}{16 \pi^2}\to 0\) and \(\frac{g^2}{\omega}\)~fixed, for arbitrary \(N_c\). The symbol '\(+\)' in the field-strength tensor \(F_{+i}\) means contraction with light-ray vector \(n_+\) and the summation over index '\(i\)' goes over two-dimensional space orthogonal to \(n_+\) and \(n_-\).
 Since the contribution of {\it\ fermions+scalars} is subleading at this limit, including the internal loops, the result is valid for the pure Yang-Mills theory as well.
  The case of two-point correlator  was elaborated in our previous paper \cite{Balitsky:2013npa} where we  defined  the generalized operators with complex spin as special light-ray operators \cite{Balitsky:1987bk} (regularized as a narrow rectangular Wilson contour called "frame") and calculated their correlator using OPE over Wilson lines \cite{Balitsky:1995ub} with a rapidity cutoff and the BFKL evolution (see \figref{fig:2p}). Here we  use the same light-ray operators: one along \(n_+\) direction and two along \(n_-\). In this case we should use more general Balitsky-Kovchegov (BK) evolution \cite{Balitsky:1997mk,Kovchegov:1999uaKovchegov:1999yj} and the leading  BFKL contribution comes from the BK vertex.

\section{Light-ray operators and their relation to local operators}
The generalization of  local operator \(\mathcal{O}^{j}\) for the case of complex spin \(j\) was constructed in  \cite{Balitsky:2013npa}. It has a form of light-ray operator \(\breve{\mathcal{S}}^J\) stretched along \(n_+\) direction and realizing the  principal series representation of \(sl(2|4)\) with conformal spin \(J=\frac{1}{2}+i\nu\) which is related to Lorentz spin \(j\)
as \(J=j+1\). The full regularized operator reads as follows:
\begin{eqnarray}
\breve{\mathcal{S}}^{j+1}(x_{1\bot})&=&\breve{\mathcal{S}}_{gl}^{j+1}(x_{1\bot})+\notag\\
+\frac{i}{2}(j-1) \breve{\mathcal{S}}_{f}^{j+1}(x_{1\bot})&-&\frac{1}{2}(j)(j-1)\breve{\mathcal{S}}_{sc}^{j+1}(x_{1\bot}),
\end{eqnarray} where for example, the regularized gluon operator is:
\begin{gather*}
\breve{\mathcal{S}}_{gl}^{j+1}(x_{1\bot})=\lim\limits_{|x_{31\bot}|\rightarrow 0}|x_{13\bot}|^{-\gamma_j} S_{gl}^{j+1}(x_{1\bot},x_{3\bot}),
\end{gather*}
\(S_{gl}^{j+1}(x_{1\bot},x_{3\bot})\!=\!\!\!\! \int\limits_{-\infty}^{\infty} \int\limits_{x_{1-}}^{\infty} \!\!\frac{d x_{1-}dx_{3-}}{x_{31-}^{j-1}} \,\ \!\!\text{tr} F^{\ \ i}_{+}(x_{1})[1,3]_{{}_\Box}F_{+i}(x_{3})\)
and \(x_1=(x_{1-},0,x_{1\bot}),\ \ x_3=(x_{3-},0,x_{3\bot})\). The anomalous dimension \(\gamma_j\) corresponds to operator \(\breve{\mathcal{S}}^{j+1}(x_{1\bot})\). Here we  introduced the notation \([1,3]_{\Box}\) for  rectangular Wilson contour  with coordinates \(x_1,x_3\) of two diagonally opposite corners, as in  \figref{fig:2p}.
  In  the case of  even integer Lorentz spin \(j\) it can be rewritten as an integral of local operator \(\mathcal{O}^{j}(x)\) with dimension \(\Delta(j)\) along a light ray direction \(n_+\):
\begin{gather}
\breve{\mathcal{S}}^{j+1}(x_{\bot})|_{j\in \text{Even}}\sim\int\limits_{-\infty}^\infty dx_- \mathcal{O}^{j}(x)
\end{gather}
In this case the correlator of two light-ray operators stretched along \(n_+\) and \(n_-\) vectors, normalized as \(\langle n_+n_-\rangle\)=1, is just the double integral of  two-point correlator of local operators w.r.t.  light-ray directions \(n_\pm\):
\begin{gather}
<\breve{\mathcal{S}}^{j_1+1}(x_{\bot}) \breve{\mathcal{S}}^{j_2+1}(y_{\bot})>=\frac{\delta(j_1-j_2) b_{j_1}}{(|x-y|^2_\bot)^{\Delta(j_1)-1}}
\end{gather}

In this note, we calculate the  correlator of three light-ray operators, restricting ourselves to a particular simple kinematics:  one light-ray operator is stretched along \(n_+\) light-ray direction and two other -- along \(n_-\). The correlator of 3 light-ray operators  can be obtained by integrating the correlator of 3 local operators along these light-rays. The tensor structures of such local correlators are known from general group-theoretical considerations  \cite{Costa:2011mg}, up to a few  structure constants depending  on the coupling and symmetry charges. The main problem which we are addressing here is the calculation of these non-trivial constants.  Remarkably, if  the coordinates of all 3 light-rays in the transverse space are restricted to the same line all these structures  collapse into a single one \cite{SimilarPhenomenon}, with a single overall structure constant which we are going to compute.
Note that after a conformal transformation the three points in the transverse space take arbitrary positions.

However, the configuration with    two collinear light-ray operators is singular, so we first consider three different light-ray directions \(n_1,\ n_2,\ n_3\)  and then take the limit \(n_2\rightarrow n_3\). The result of integration along light-rays is quite simple and contains only one unknown overall constant
\begin{gather}
\langle\breve{\mathcal{S}}^{j_1+1}(x_{\bot})\breve{\mathcal{S}}^{j_2+1}(y_{\bot})\breve{\mathcal{S}}^{j_3+1}(z_{\bot})\rangle=C_{\{n_i\}}(\{\Delta_i\},\{j_i\})\cdot\notag\\
\cdot\frac{<n_1n_2>^{[j]_{1,2;3}}<n_1n_3>^{[j]_{1,3;2}}<n_2n_3>^{[j]_{2,3;1}}}
{(|x-y|_{\bot}^2)^{[\Delta]_{1,2;3}}(|x-z|_{\bot}^2)^{[\Delta]_{1,3;2}}(|y-z|_{\bot}^2)^{[\Delta]_{2,3;1}}}
\end{gather}
where we  used a short-hand notation \([a]_{i,j;k}\equiv \frac{1}{2}(a_i+a_j-a_k-1)\) and \(\{a_i\}\equiv\{a_1,a_2,a_3\}\).
In what follows, we assume  the existence of a good analytic continuation for \(C_{\{n_i\}}(\{\Delta(j_i)\},\{j_i\})\) to non-integer \(\{j_i\}\)'s. We take the limit \(n_1=n_+,\ n_2=n_-,\ n_3\rightarrow n_2\) with the normalization \(\langle n_+ n_-\rangle=1\). In BFKL regime \(j_i=1+\omega_i\rightarrow 1\) we obtain:
\begin{gather}
\langle\breve{\mathcal{S}}^{2+\omega_1}(x_{\bot})\breve{\mathcal{S}}^{2+\omega_2}(y_{\bot})\breve{\mathcal{S}}^{2+\omega_3}(z_{\bot})\rangle=\notag\\
=\lim\limits_{n_3\rightarrow n_2=n_-} \frac{<n_2n_3>^{\frac{\omega_2+\omega_3-\omega_1}{2}}}{\omega_2+\omega_3-\omega_1}\times\notag\\
\times \frac{C_{+--}(\{\Delta_i\},\{1+\omega_i\})}
{|x-y|_{\bot}^2)^{[\Delta]_{1,2;3}}(|x-z|_{\bot}^2)^{[\Delta]_{1,3;2}}(|y-z|_{\bot}^2)^{[\Delta]_{2,3;1}}}\label{3pFromLocals}
\end{gather}
where \(\Delta_i=\Delta(1+\omega_i,g^2) \)  is given by BFKL spectrum (see below).
We  explicitly pulled out   the denominator \(\frac{1}{\omega_2+\omega_3-\omega_1}\) because it will emerge in our forthcoming calculation using the BK evolution. We interpret  \(\lim\limits_{<n_2n_3>\rightarrow 0}\frac{<n_2n_3>^{\frac{\omega_2+\omega_3-\omega_1}{2}}}{\omega_2+\omega_3-\omega_1}\) as a delta function \(\delta(\omega_2+\omega_3-\omega_1)\) reflecting the boost invariance. In addition, we keep  \(\omega_i\) positive through the paper.

Finally, the  structure constant is normalized using the corresponding 2-point correlators:
\begin{gather}\label{normC}
C_{\omega_1,\omega_2,\omega_3}=\frac{C_{+--}
(\{\Delta_i\},\{1+\omega_i\})}{\sqrt{b_{1+\omega_1}b_{1+\omega_2}b_{1+\omega_3}}}
\end{gather}

\section{Decomposition over dipoles and BK evolution}
When calculating the two-point correlator  \cite{Balitsky:2013npa} we  used a point splitting regularization in orthogonal direction, replacing light-rays by infinitely narrow Wilson frames with inserted fields in the corners (see \figref{fig:2p}). Now, for the  sake of simplicity, we carry out our calculation for pure Wilson frames, related to our operators with zero \(R\)-charge in the following way:
\begin{gather}
\partial_{x_{1\bot }}\cdot\partial_{x_{3\bot}} \int \int\frac{dx_{1-}dx_{3-}}{(x_{3-}-x_{1-})^{2+\omega_1}}
\,\,[x_1,x_3]_{\Box}\rightarrow \notag\\
\underset{x_{13\bot}\rightarrow 0,\,\omega_1 \rightarrow 0}{\rightarrow} |x_{13\bot}|^{\gamma_{j_1}}c(g_{YM}^2,N_c,\omega_1)\breve{\mathcal{S}}^{2+\omega_1}(x_{1\bot}).\label{LightRayFromRamka}
\end{gather}
The   coefficient \(c(g_{YM}^2,N_c,\omega_i)\) (denoted below as \(c(\omega_i))\)  depends on the local regularization procedure and at weak coupling it behaves as \(c(\omega_i)\sim\frac{g^2_{YM}}{\omega_i}\), but its explicit form is irrelevant for us because we are going to calculate the normalized structure constant where it cancels.   In general, there are a few types of leading twist-2 operators which appear in this decomposition  but in the BFKL limit a single one with the smallest anomalous dimension survives. In addition, in the \(\omega_i\to 0 \) limit only the term  built out of gauge fields alone does contribute~\cite{Balitsky:2013npa}.

Following the OPE method \cite{Balitsky:1995ub},       the pure Wilson frames can be replaced by regularized color dipoles:
\begin{gather}
[x_1,x_3]_{\Box} \rightarrow N(1-\mathbf{U}^{\sigma_+}(x_{1\bot},x_{3\bot}))
\end{gather}
where
\begin{gather}
\mathbf{U}^{\sigma_+}(x_{1\bot},x_{3\bot})=1-\frac{1}{N}\text{tr}(U^{\sigma_+}_{x_{1\bot}}U^{\sigma_+ \dagger}_{x_{3\bot}}),\\
U^{\sigma_+}_{x_\bot}=P\exp[i g_{{}_{YM}}
\int\limits_{-\infty}^\infty dx_+A^{\sigma_+}_-(x)],\\
A_{\mu}^{\sigma_+}(x)=\int d^4k \theta(\sigma_+-|k_+|)e^{ikx}A_\mu(k)
\end{gather}
and \(\sigma_+ \)  is a longitudinal cutoff in \(n_+\) direction. Now we can write:
\begin{gather}
\langle S^{2+\omega_1}(x_{1\bot},x_{3\bot})S^{2+\omega_2}(y_{1\bot},y_{3\bot})S^{2+\omega_3}(z_{1\bot},z_{3\bot})\rangle=\notag\\
={\cal -D}_{\bot}\int\limits_{-\infty}^{\infty}dx_{1-} \int\limits_{x_{1-}}^{\infty}dx_{3-}x_{31-}^{-2-\omega_1}\times\label{3pStart}\\ \times \int\limits_{-\infty}^{\infty}dy_{1+} \int\limits_{y_{1+}}^{\infty}dy_{3+}y_{31+}^{-2-\omega_2}
 \int\limits_{-\infty}^{\infty}dz_{1+} \int\limits_{z_{1+}}^{\infty}dz_{3+}z_{31+}^{-2-\omega_3}\times \notag \\
\times\langle \mathbf{U}^{\sigma_{1-}}(x_{1\bot},x_{3\bot})\mathbf{V}^{\sigma_{2+}}(y_{1\bot},y_{3\bot})\mathbf{W}^{\sigma_{3+}}(z_{1\bot},z_{3\bot}) \rangle,\notag
\end{gather}
where \({\cal D}_{\bot}=\frac{N^3(\partial_{x_{1\bot}}\cdot\partial_{x_{3\bot}})(\partial_{y_{1\bot}}
\cdot\partial_{y_{3\bot}})(
\partial_{z_{1\bot}}\cdot\partial_{z_{3\bot}})}{c(\omega_1)c(\omega_2)c(\omega_3)}\).

In our kinematics two dipoles \(\mathbf{V}\) and \(\mathbf{W}\) have zero \(n_+\) projection and in the  BFKL approximation they form a "pancake" field configuration in the reference frame related to \(\mathbf{U}\). This means that the rapidity of \(\mathbf{U}\) serves as the upper limit for  integrations w.r.t. rapidities of     \(\mathbf{V}\) and \(\mathbf{W}\)  in our logarithmic approximation. Now we use the BK evolution equation \cite{Balitsky:1997mk,Kovchegov:1999uaKovchegov:1999yj} to calculate the quantum average in (\ref{3pStart}). It gives the evolution of the  dipole \(\mathbf{U}^{Y}\) with respect to rapidity \(Y=e^{\sigma}\), namely
\begin{gather}
\sigma \frac{d}{d\sigma}\mathbf{U}^\sigma(z_1,z_2)=\mathcal{K}_{{\rm BK}}\ast \mathbf{U}^\sigma(z_1,z_2), \label{BK}
\end{gather}
where \(\mathcal{K}_{\rm BK}\) is an integral operator having the following form in LO  approximation:
\begin{gather}
\mathcal{K}_{_{\rm LO\, BK}}\ast\mathbf{U}(z_1,z_2)=\frac{2g^2}{\pi} \int d^2z_3\frac{z_{12}^2}{z_{13}^2z_{23}^2}\left[\mathbf{U}(z_1,z_3)+\right.\notag\\
\left. +\mathbf{U}(z_3,z_2)-\mathbf{U}(z_1,z_2)-\mathbf{U}(z_1,z_3)\mathbf{U}(z_3,z_2)\right].
\end{gather}
Evolution of \(\mathbf{U}^{Y_1}\) goes from \(Y_1\) to an intermediate \(Y_0\) w.r.t. the linear part of (\ref{BK}), and then the BK vertex   acts at  \(Y_0\) and generates two dipoles which can be contracted with \(\mathbf{V}^{Y_{2}}\) and \(\mathbf{W}^{Y_{3}}\). Schematically, it can be written as:
\begin{gather*}
\int dY_0(\mathbf{U}^{Y_{1}}\rightarrow \mathbf{U}^{Y_0})\otimes(\text{BK\ vertex at}\,\,Y_0 )\otimes\begin{pmatrix}\langle \mathbf{U}^{Y_{0}} \mathbf{V}^{Y_{2}}\rangle \\
\langle \mathbf{U}^{Y_{0}} \mathbf{W}^{Y_{3}}\rangle \\
\end{pmatrix}
\end{gather*}
The linear BFKL evolution of \(\mathbf{U}^{Y_1}\) from \(Y_1\) to \(Y_0\) gives:
\begin{gather}
\mathbf{U}^{Y_1}(x_1,x_3)=\int d\nu\int d^2x_0 \frac{\nu_1^2}{\pi^2}E_{\nu_1}(x_{10},x_{30})e^{\aleph(\nu_1)Y_{10}}\cdot\notag\\
\cdot\frac{1}{\pi^2}\int
\frac{d^2\gamma d^2\beta}{|\gamma-\beta|^4}E^{*}_{\nu_1}(\gamma-x_0,\beta-x_0)\mathbf{U}^{Y_0}(\gamma,\beta),
\end{gather}
where we denoted \(Y_{ij}\equiv Y_i-Y_j\) and we  introduced the function \(E_\nu(z_{10},z_{20})=(\frac{|z_{12}|^2}{|z_{10}|^2|z_{20}|^2})^{1/2+i\nu}\) which projects dipoles on the eigenstates of BFKL operator with the eigenvalues \(\aleph(\nu)=4g^2(2\psi(1)-\psi(1/2+i\nu)-\psi(1/2-i\nu))\).    We take here only the sector  \(n=0\), where \(n\) is the discrete quantum number of \(SL(2,C)\) because it gives the leading contribution.

The non-linear part of BK evolution   (\ref{BK}) is described by the following renorm group equation:
\begin{gather}
\left.\frac{\partial}{\partial Y}\mathbf{U}^{Y}(\gamma,\beta)\right|_{Y=Y_0}=\notag\\-\frac{2g^2}{\pi}
 \int d^2\alpha\frac{|\gamma-\beta|^2}{|\gamma-\alpha|^2|\beta-\alpha|^2}\mathbf{U}^{Y_0}(\gamma,\alpha)\mathbf{U}^{Y_0}(\alpha,\beta)
\end{gather}
Finally, we  contract the two emerging dipoles \(\mathbf{U}^{Y_0}(\gamma,\alpha)\) and \(\mathbf{U}^{Y_0}(\alpha,\beta)\) with \(\mathbf{V}^{\sigma_{2+}}(y_{1\bot},y_{3\bot})\) and \(\mathbf{W}^{\sigma_{3+}}(z_{1\bot},z_{3\bot})\). Thus for the planar contribution we get:
\begin{gather}
\langle \mathbf{U}^{Y_{1}}(x_{1\bot},x_{3\bot})\mathbf{V}^{Y_{2}}(y_{1\bot},y_{3\bot})\mathbf{W}^{Y_{3}}(z_{1\bot},z_{3\bot}) \rangle_{pl}=\label{evolutionUpDown}\\
=-\frac{2g^2}{\pi}\int d Y_0 \int d\nu_1\int d^2x_0 \frac{\nu_1^2}{\pi^2}E_{\nu_1}(x_{10},x_{30})e^{\aleph(\nu_1)Y_{10}}\times\notag\\
\times\frac{1}{\pi^2}\int
\frac{d^2\alpha d^2\beta d^2\gamma}{|\gamma-\beta|^2 |\gamma-\alpha|^2|\beta-\alpha|^2}E^{*}_{\nu_1}(\gamma-x_0,\beta-x_0)\cdot \notag\\
\cdot  (\langle \mathbf{U}^{Y_0}(\gamma,\alpha)\mathbf{V}^{Y_{2}}(y_{1\bot},y_{3\bot})\rangle \langle\mathbf{U}^{Y_0}(\alpha,\beta) \mathbf{W}^{Y_{3}}(z_{1\bot},z_{3\bot}) \rangle +\notag\\
+\langle \mathbf{U}^{Y_0}(\gamma,\alpha)\mathbf{W}^{Y_{3}}(z_{1\bot},z_{3\bot})\rangle \langle\mathbf{U}^{Y_0}(\alpha,\beta) \mathbf{V}^{Y_{2}}(y_{1\bot},y_{3\bot}) \rangle)\notag
\end{gather}
The last two terms in (\ref{evolutionUpDown}) give the same contribution so it is enough  to know the correlators of two dipoles \cite{Balitsky:2013npa}:
\begin{gather}
\langle \mathbf{U}^{Y_0}(\gamma,\alpha)\mathbf{V}^{Y_{2}}(y_{1\bot},y_{3\bot})\rangle=\frac{8g^4(1-N_c^2)}{N_c^4}\int d^2y_0 \cdot \notag \\
\cdot \int\frac{d\nu_2 \nu_2^2 e^{Y_{02}\aleph(\nu_2)}}{(\frac{1}{4}+\nu_2^2)^2} E_{\nu_2}(\gamma-y_0,\alpha-y_0) E^*_{\nu_2}(y_{10},y_{30}) \label{NizhnyProp2}
\end{gather}
and similarly for \(\langle\mathbf{U}^{Y_0}(\alpha,\beta) \mathbf{W}^{Y_{3}}(z_{1\bot},z_{5\bot}) \rangle\).
It was argued in  \cite{Balitsky:2013npa} that we can make the following identification for rapidities in dipole correlators:
 \( Y_{02}=\ln \frac{L_0y_{31+}}{\Lambda^2}, \ Y_{03}=\ln\frac{L_0z_{31+}}{\Lambda^2}\), where \(\Lambda\) a cutoff whose precise value is irrelevant in LO.
On the other hand, the difference of rapidities of the first dipole and of the BK vertex  \( Y_{10}=\ln \frac{x_{31-}}{L_0} \) corresponds to BFKL evolution.
The integral over  \(Y_0=\ln\frac{L_0}{\Lambda}\) goes from \(Y_1\) to \(\text{max}(Y_2,Y_3)\).
If we plug (\ref{evolutionUpDown})-(\ref{NizhnyProp2}) into (\ref{3pStart}) and do the integrals over light ray directions, i.e. over rapidities, we obtain the following planar contribution:
\begin{gather}
\langle S^{2+\omega_1}(x_{1\bot},x_{3\bot})S^{2+\omega_2}(y_{1\bot},y_{3\bot})S^{2+\omega_3}(z_{1\bot},z_{3\bot})\rangle_{pl}=\notag\\
=\frac{2^8 g^{10}(N_c^2-1)^2}{\pi^3 N_c^8}\delta(\omega_1-\omega_2-\omega_3)D_{\bot}\notag\\
\int d\nu_1\frac{\nu_1^2}{\pi^2}\frac{1}{\omega_2+\omega_3-\aleph(\nu_1)} \int \frac{d\nu_2 \nu_2^2}{(\frac{1}{4}+\nu_2^2)^2} \frac{1}{\omega_2-\aleph(\nu_2)} \cdot\notag \\ \cdot \int  \frac{d\nu_3 \nu_3^2}{(\frac{1}{4}+\nu_3^2)^2}\frac{1}{\omega_3-\aleph(\nu_3)} \int d^2x_0 d^2y_0 d^2z_0  E^*_{\nu_1}(x_{10},x_{30})\cdot\notag\\
\cdot E^*_{\nu_2}(y_{10},y_{30}) E^*_{\nu_3}(z_{10},z_{30})\Upsilon_{pl} (\nu_1,\nu_2,\nu_3;x_0,y_0,z_0) \label{Ramki3pConfRepPlanar}
\end{gather}
The usual delta-function $\delta(\omega_1-\omega_2-\omega_3)$ (see e.g. \cite{Bartels:1994jj})
is a consequence of boost-invariance as in the formula (\ref{3pFromLocals}).  \(\Upsilon_{pl}\) represents the planar contribution of BK vertex:
\begin{gather}
\Upsilon_{pl} (\nu_1,\nu_2,\nu_3;x_0,y_0,z_0)=\notag\\
=\int\frac{d^2\alpha d^2\beta d^2\gamma}{|\gamma-\beta|^2 |\gamma-\alpha|^2|\beta-\alpha|^2}
E_{\nu_1}(\beta-x_0,\gamma-x_0)\cdot \notag\\
\cdot E_{\nu_2}(\alpha-y_0,\gamma-y_0)E_{\nu_3}(\alpha-z_0,\beta-z_0)=\label{Upsilon_pl}\\
=\frac{\Omega(h_1,h_2,h_3)}
{|x_0-y_0|^{4[h]_{1,2;3}+2}\ |x_0-z_0|^{4[h]_{1,3;2}+2}\ |y_0-z_0|^{4[h]_{2,3;1}+2}} \notag
\end{gather}
where \(h_1=\frac{1}{2}+i\nu_1,\,h_2=\frac{1}{2}+i\nu_2,\,h_3=\frac{1}{2}+i\nu_3\) and the function \(\Omega(h_1,h_2,h_3)\) was presented in  \cite{Korchemsky:1997fy}.

\begin{figure}
  \centering
  \includegraphics[scale=0.2]{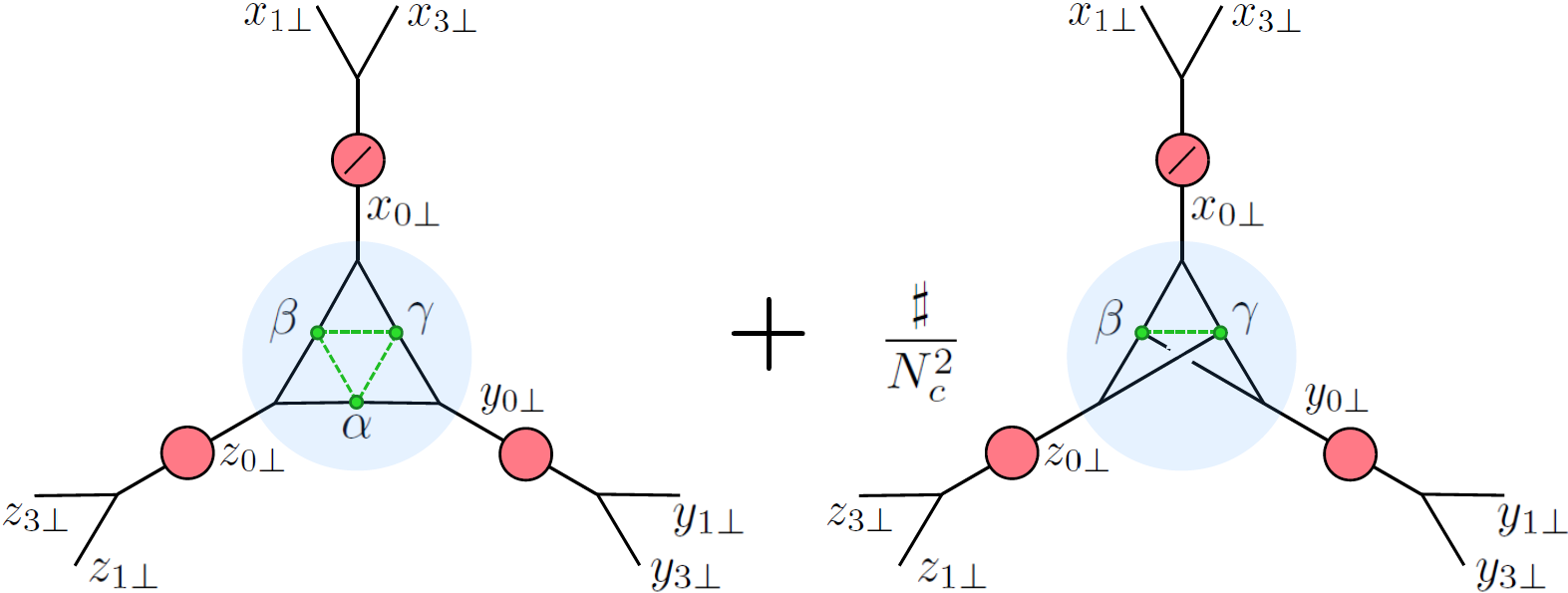}
  \caption{The structure of 3-point correlator. Red circles correspond to BFKL propagators (the crossed one has extra multiplier \((\frac{1}{4}+\nu_1^2)^2\)). The blue blob  corresponds to the  3-point functions of 2-dimensional  BFKL CFT.  The triple "Y"-veritces correspond to \(E\)-functions. For example vertex with ends labeled as \(z_{1\bot},z_{3\bot},z_{0\bot}\) corresponds to \(E_\nu(z_{10\bot},z_{30\bot})=(\frac{|z_{13\bot}|^2}{|z_{10\bot}|^2|z_{30\bot}|^2})^{1/2+i\nu_3}\). The \(\alpha\beta\gamma\)-triangle in the first, planar, term and \(\beta\gamma\)-link in the second, nonplanar, term correspond to  triple pomeron vertex.}
  \label{fig:3p}
\end{figure}

Remarkably, we can also take into account the non-planar contribution \cite{Korchemsky:1997fy,Chirilli:2010mw},   thus providing the finite \(N_c\) answer for the BFKL structure constant! It appears as a single extra term \(\Upsilon_{npl}\):
\begin{gather}
\Upsilon_{npl} (\nu_1,\nu_2,\nu_3;x_0,y_0,z_0)
=\int\frac{d^2\beta d^2\gamma}{|\gamma-\beta|^4} E_{\nu_1}(\beta-x_0, \notag\\
\gamma-x_0)E_{\nu_2}(\beta-y_0,\gamma-y_0)E_{\nu_3}(\beta-z_0,\gamma-z_0)=\label{Upsilon_npl}\\
=\frac{\Lambda(h_1,h_2,h_3)}
{|x_0-y_0|^{4[h]_{1,2;3}+2}\ |x_0-z_0|^{4[h]_{1,3;2}+2}\ |y_0-z_0|^{4[h]_{2,3;1}+2}} \notag
\end{gather}
where  \(\Lambda(h_1,h_2,h_3)\) was also presented in  \cite{Korchemsky:1997fy}, and the full answer can be obtained from (\ref{Ramki3pConfRepPlanar}) by replacing \(\Upsilon_{pl}\) with \(\Upsilon\) (see in \figref{fig:3p}):
\begin{gather}
\Upsilon=\Upsilon_{pl}-\frac{2\pi}{N^2}\Upsilon_{npl} Re [ \psi(1)+\notag\\
+\psi(\frac{1}{2}+i\nu_1)-\psi(\frac{1}{2}+i\nu_2)-\psi(\frac{1}{2}+i\nu_3)].\label{Upsilon}
\end{gather}
The integrals over \(x_0,y_0,z_0\) are easily computable, e.g.
\begin{gather}
\int d^2x_0 E_{\nu_1}(\beta-x_0,\gamma-x_0)E^*_{\nu_1}(x_{10},x_{30})=\notag\\
=(\tau^2)^{\frac{1}{2}+i\nu_1}
{}_2F_1(\frac{1}{2}+i\nu,\frac{1}{2}+i\nu,1+2i\nu,\tau)\times\notag\\
\times {}_2F_1(\frac{1}{2}+i\nu,\frac{1}{2}+i\nu,1+2i\nu,\bar{\tau})\frac{(\frac{1}{4}+\nu^2)^2}{\nu^2}G(\nu)+\notag\\
+(\nu\leftrightarrow -\nu),\label{Lipatov4p}\\
G(\nu)=\frac{\nu^2}{(\frac{1}{4}+\nu^2)^2}\frac{\pi \Gamma^2(\frac{1}{2}+i\nu)\Gamma(-2i\nu)}{\Gamma^2(\frac{1}{2}-i\nu)\Gamma(1+2i\nu)},
\end{gather}
where \(\tau=\frac{|x_1-x_3||\beta-\gamma|}{|x_1-\beta||x_3-\gamma|}\). In the limit \(x_1,x_3\rightarrow x\) we can replace \(\frac{|x_1-x_3||\beta-\gamma|}{|x_1-\beta||x_3-\gamma|}\rightarrow\frac{|x_1-x_3||\beta-\gamma|}{|x-\beta||x-\gamma|}\to 0\). For small \(\tau\) we  close the  \(\nu_1\) contour in the lower (upper) half-plane for first(second) term, respectively, both of them giving the same contribution. Integrals over \(\alpha,\beta,\gamma\)  in \eqref{Ramki3pConfRepPlanar} can be  reduced to   \(\Upsilon_{pl}\)   represented in \cite{Korchemsky:1997fy} in terms of   hypergeometric and Meijer G functions, and \(\Upsilon_{npl}\)   in terms of \(\Gamma\)-functions. Integrals over \(\nu_i\) can be done by picking up the BFKL poles \(\omega_i=\aleph(\nu^*_i)\).

Combining (\ref{Ramki3pConfRepPlanar}),(\ref{Upsilon}) and (\ref{Lipatov4p}) we come to the final expression for 3-point correlation function:
\begin{gather}
\langle S^{2+\omega_1}(x_{1\bot},x_{3\bot})S^{2+\omega_2}(y_{1\bot},y_{3\bot})S^{2+\omega_3}(z_{1\bot},z_{3\bot})\rangle=\notag\\
=-i g^{10} \frac{\delta(\omega_1-\omega_2-\omega_3)}{c(\omega_1)c(\omega_2)c(\omega_3)}H\cdot\\
\cdot\frac{\Psi(\nu_1^*,\nu_2^*,\nu_3^*)|x_{13}|^{\gamma_1}|y_{13}|^{\gamma_2}|z_{13}|^{\gamma_3}}
{|x-y|^{2+\gamma_1+\gamma_2-\gamma_3}|x-z|^{2+\gamma_1+\gamma_3-\gamma_2}|y-z|^{2+\gamma_2+\gamma_3-\gamma_1}}\label{Otvet}\notag
\end{gather}
\begin{gather}
\text{where}\quad H=\frac{2^{10}(N_c^2-1)^2}{\pi^2 N_c^5}\gamma_1^2(2+\gamma_1)^4(2+\gamma_2)^2\times\notag\\
\times(2+\gamma_3)^2\frac{G(\nu_1^*)}{\aleph'(\nu_1^*)} \frac{G(\nu_2^*)}{\aleph'(\nu_2^*)} \frac{G(\nu_3^*)}{\aleph'(\nu_3^*)},
\label{3pointfinal}\end{gather}
\(\gamma_i=\gamma(1+\omega_i)\) - anomalous dimension and the coefficient \(\Psi(\nu_1^*,\nu_2^*,\nu_3^*)\) can be expressed through the  functions \(\Omega(h_1,h_2,h_3)\)  and \(\Lambda(h_1,h_2,h_3)\) defined in (\ref{Upsilon_pl})-(\ref{Upsilon_npl}) and calculated in \cite{Korchemsky:1997fy}:
\begin{gather}
\Psi(\nu_1^*,\nu_2^*,\nu_3^*)=\Omega(h^*_1,h^*_2,h^*_3)-\frac{2\pi}{N_c^2}\Lambda(h^*_1,h^*_2,h^*_3)\cdot\notag\\
\cdot\text{Re}(\psi(1)-\psi(h^*_1)-\psi(h^*_2)-\psi(h^*_3)),
\end{gather}
where \(h^*_i=\frac{1}{2}+i\nu_i^*=1+\frac{\gamma_i}{2}\).

Our final result for  normalized structure constant   is:
\begin{gather}
C_{\omega_1,\omega_2,\omega_3}=-i^{1/2}g^4 \frac{2}{\pi^5}\frac{\sqrt{N_c^2-1}}{N_c^2}\gamma_1^2(2+\gamma_1)^2\cdot\notag\\
\cdot\sqrt{\frac{G(\nu_1^*)}{\aleph'(\nu_1^*)}\frac{G(\nu_2^*)}{\aleph'(\nu_2^*)}\frac{G(\nu_3^*)}{\aleph'(\nu_3^*)}}
\Psi(\nu_1^*,\nu_2^*,\nu_3^*)\label{FinalResult}
\end{gather}
Precising the dependence on parameters \(\{\frac{g^2}{\omega_i}\}\), \(g^2\) and \(N_c\) we can write: \(C_{\omega_1,\omega_2,\omega_3}=g \frac{\sqrt{N_c^2-1}}{N_c^2} f(\frac{g^2}{\omega_1},\frac{g^2}{\omega_2},\frac{g^2}{\omega_3})\),
where \(f\) is a function which depends only on  the ratios \(\{\frac{g^2}{\omega_i}\}\). In the limit \(\frac{g^2}{\omega_i}\rightarrow 0\) we get the  asymptotics:
\begin{gather}
\Omega(h_1^*,h_2^*,h_3^*)\rightarrow -\frac{16 \pi^3}{\gamma_1^2\gamma_2^2\gamma_3^2}
\cdot[\gamma_1^2(\gamma_2
+\gamma_3)
+\gamma_2^2(\gamma_1+\gamma_3)+\notag\\+\gamma_3^2(\gamma_1+\gamma_2)
+\gamma_1 \gamma_2 \gamma_3)(1+O(g^2/\omega_i))\notag\\
\Lambda(h_1^*,h_2^*,h_3^*)\rightarrow \frac{8\pi^2 (\gamma_1+\gamma_2+\gamma_3)}{\gamma_1 \gamma_2 \gamma_3}(1+O(g^2/\omega_i))
\end{gather}
In this limit \(\gamma_i=-\frac{8g^2}{\omega_i}+o(\frac{g^2}{\omega_i})\) and the main contribution to 3-point correlator  \eqref{FinalResult} comes from the planar \(\mathcal{O}(g^2)\) term
\begin{gather}
C_{\omega_1,\omega_2,\omega_3}=-ig^2 \frac{\sqrt{N_c^2-1}}{\sqrt{2\pi}N_c^2} \frac{1}{\omega_1^\frac{5}{2}\omega_2^\frac{1}{2}\omega_3^\frac{1}{2}}(\omega_1^2(\omega_2+\omega_3)+\notag\\
\omega_2^2(\omega_1+\omega_3)+\omega_3^2(\omega_1+\omega_2)+\omega_1 \omega_2 \omega_3)(1+O(g^2)) \label{Cpert2}
\end{gather}
whereas the nonplanar one is \(\mathcal{O}(g^6)\). It might seem strange that the planar contribution does not start from  \(\mathcal{O}(g^4)\) terms given by the leading Feynman graphs, e.g.  with 4 gluon vertices. However, in BFKL approximation we   should keep \(\frac{g^2}{\omega}\gg\omega\) \cite{Simon}. In addition, when making the point-splitting regularization we have to keep \(\frac{g^2}{\omega}|\ln (x_{31\bot}/(x-y))^2|\gg 1\). The limit \(|x_{13\bot}|\) has to be taken first, which makes the value \(g^2=0\) exceptional.  This order of limits leads to \({\cal\ O}(g^2)\) behavior of \eqref{Cpert2}.

\section{Discussion}
Our result eq.(\ref{FinalResult}), based on BFKL approximation  is a rare example of computation of a non-BPS structure constant receiving contributions from all orders in coupling constant, including infinitely many "wrapping" corrections.  Moreover, our result is valid at any \(N_c\). Since in the LO BFKL the contributions of all  fields but gluons  in \(\mathcal{N}=4 \) SYM disappear from both the definition of operators and  internal loops, the result is applicable to pure YM theory at any \(N_c\), including \(N_c=3\).  It would be interesting to apply our structure constants to the OPE at hard scattering in real QCD and to work out the full "dictionary" relating them to the OPE in the 2-dimensional \(SL(2,C)\) CFT -- the basis of our BFKL computation.   It is also not hopeless, though challenging, to compute these structure constants in the NLO approximation in \(\mathcal{N}=4 \) SYM. Our present result may serve as an important, all-wrappings test for the future computations of similar quantities in the integrability approaches to planar AdS\(_5\)/CFT\(_4\), such as          \cite{Basso:2015zoa} and the BFKL limit of quantum spectral curve \cite{Alfimov:2014bwa}.
\begin{acknowledgments}
\section*{Acknowledgments}
\label{sec:acknowledgments}
We thank J.~Bartels, S.~Caron-Huot, L.~Lipatov and V.~Schomerus for  discussions. Our special thanks  to G.~Korchemsky who participated in the initial stage of this work. The work of E.S. and V.K. was supported by the People Programme (Marie Curie Actions) of the European Union's Seventh Framework Programme FP7/2007-2013/ under REA Grant Agreement No 317089 (GATIS).
The work of
V.K.  has received funding from the European Research Council (Programme
”Ideas” ERC-2012-AdG 320769 ”AdS-CFT-solvable”), from the ANR grant StrongInt (BLANC- SIMI-
4-2011) and from the ESF grant HOLOGRAV-09- RNP- 092.
The work of I.B.  was supported by DOE contract
 DE-AC05-06OR23177  and by the grant DE-FG02-97ER41028.

\end{acknowledgments}


\begin{thebibliography}{99}%

\bibitem{Ioffe:2010zz}
  B.~L.~Ioffe, V.~S.~Fadin and L.~N.~Lipatov,
  ``Quantum chromodynamics: Perturbative and nonperturbative aspects,''

\bibitem{Kovchegov:2012mbw}
  Y.~V.~Kovchegov and E.~Levin,
  ``Quantum chromodynamics at high energy,''

\bibitem{Beisert:2010jrGromov:2014caa}
  N.~Beisert, C.~Ahn, L.~F.~Alday, Z.~Bajnok, J.~M.~Drummond, L.~Freyhult, N.~Gromov and R.~A.~Janik {\it et al.},
  Lett.\ Math.\ Phys.\  {\bf 99} (2012) 3
  [arXiv:1012.3982 [hep-th]];
  N.~Gromov, V.~Kazakov, S.~Leurent and D.~Volin,
  arXiv:1405.4857 [hep-th].

\bibitem{Basso:2015zoa}
  B.~Basso, S.~Komatsu and P.~Vieira,
  arXiv:1505.06745 [hep-th].

\bibitem{Fadin:1975cbBalitsky:1978ic}
V.~S.~Fadin, E.~A.~Kuraev and L.~N.~Lipatov,
  Phys.\ Lett.\ B {\bf 60} (1975) 50;
I.~Balitsky and L.~Lipatov,
    {\em Sov.J.Nucl.Phys.} {\bf 28} (1978) 822--829.
\bibitem{Balitsky:2013npa}
  I.~Balitsky, V.~Kazakov and E.~Sobko,
  arXiv:1310.3752 [hep-th].

\bibitem{Balitsky:1987bk}
  I.~Balitsky and V.~M.~Braun,
  Nucl.\ Phys.\ B {\bf 311} (1989) 541.

\bibitem{Balitsky:1995ub}
  I.~Balitsky,
  Nucl.\ Phys.\ B {\bf 463} (1996) 99
  [hep-ph/9509348].

\bibitem{Balitsky:1997mk}
  I.~Balitsky,
  AIP Conf.\ Proc.\  {\bf 407} (1997) 953
  [hep-ph/9706411].

\bibitem{Kovchegov:1999uaKovchegov:1999yj}
  Y.~V.~Kovchegov,
  Phys.\ Rev.\ D {\bf 60} (1999) 034008
  [hep-ph/9901281];
  Y.~V.~Kovchegov,
  Phys.\ Rev.\ D {\bf 61} (2000) 074018
  [hep-ph/9905214].

\bibitem{Costa:2011mg}
  M.~S.~Costa, J.~Penedones, D.~Poland and S.~Rychkov,
  JHEP {\bf 1111} (2011) 071
  [arXiv:1107.3554 [hep-th]].

\bibitem{Kazakov:2012ar}
  V.~Kazakov and E.~Sobko,
  JHEP {\bf 1306} (2013) 061
  [arXiv:1212.6563 [hep-th]].

\bibitem{SimilarPhenomenon}
The similar phenomenon was observed in \cite{Kazakov:2012ar}

\bibitem{Bartels:1994jj}
  J.~Bartels and M.~Wusthoff,
  Z.\ Phys.\ C {\bf 66} (1995) 157.

\bibitem{Korchemsky:1997fy}
  G.~P.~Korchemsky,
  Nucl.\ Phys.\ B {\bf 550} (1999) 397
  [hep-ph/9711277].

\bibitem{Chirilli:2010mw}
  G.~A.~Chirilli, L.~Szymanowski and S.~Wallon,
  Phys.\ Rev.\ D {\bf 83} (2011) 014020
  doi:10.1103/PhysRevD.83.014020
  [arXiv:1010.0285 [hep-ph]].

\bibitem{Simon}
We thank S.~Caron-Huot for the discussion on this subject.

\bibitem{Alfimov:2014bwa}
  M.~Alfimov, N.~Gromov and V.~Kazakov,
  arXiv:1408.2530 [hep-th].


\end{thebibliography}

%

\end{document}